\def\withFigures{1}  % if we want figures included, replace 0 -> 1
\documentclass[a4paper,11pt]{article}
\pdfoutput=1
\usepackage{jheppub}
\usepackage{epsfig,amsfonts,verbatim,mathrsfs}
\usepackage{dcolumn,bm,multirow}
\newcommand{\beq}{\begin{eqnarray}}
\newcommand{\eeq}{\end{eqnarray}}
\newcommand{\bea}{\begin{eqnarray}}
\newcommand{\eea}{\end{eqnarray}}
\newcommand{\lrf}[2]{\left(\frac{#1}{#2}\right)}

\newcommand{\stp}{\tilde{t}}
\newcommand{\nnmb}{\nonumber}
\newcommand{\del}{\partial}
\newcommand{\tev}{\, {\rm TeV}}
\newcommand{\gev}{\, {\rm GeV}}

\newcommand{\drbar}{\overline{\mathrm{DR}}}
\renewcommand{\vec}[1]{\boldsymbol{\mathbf{#1}}}
\newcommand{\bvec}[1]{\bar{\vec{#1}}}

\title{Vacuum Stability and the MSSM Higgs Mass}
\author[a,b]{Nikita Blinov}
\author[a]{David E. Morrissey}
\affiliation[a]{TRIUMF, 4004 Wesbrook Mall, Vancouver, BC V6T~2A3, Canada}
\affiliation[b]{Department of Physics and Astronomy, University of British Columbia,\\ 
Vancouver, BC V6T~1Z1, Canada}
\emailAdd{nblinov@triumf.ca}
\emailAdd{dmorri@triumf.ca}
\abstract{
In the Minimal Supersymmetric Standard Model~(MSSM), 
a Higgs boson mass of 125 GeV can be obtained with moderately
heavy scalar top superpartners provided they are highly mixed.
The source of this mixing, a soft trilinear stop-stop-Higgs coupling, 
can result in the appearance of charge- and color-breaking minima in 
the scalar potential of the theory.  If such a vacuum exists and
is energetically favorable, the Standard Model-like vacuum can decay 
to it via quantum tunnelling.  In this work we investigate the conditions
under which such exotic vacua arise, and we compute the tunnelling
rates to them.  Our results provide new constraints on the scalar top
quarks of the MSSM.
}
\arxivnumber{1310.4174}
\begin{document}
\maketitle

\begin{comment}
\begin{titlepage}
\noindent
\vspace{1cm}
%
\begin{center}
  \begin{Large}
    \begin{bf}
Vacuum Stability and the MSSM Higgs Mass
     \end{bf}
  \end{Large}
\end{center}
%
\vspace{0.2cm}

\begin{center}

\begin{large}
Nikita Blinov$^{(a,b)}$ and
David E. Morrissey$^{(a)}$
\end{large}
\vspace{1cm}\\
%
  \begin{it}
%\begin{flushleft}
(a) TRIUMF, 4004 Wesbrook Mall, Vancouver, BC V6T~2A3, Canada\vspace{0.3cm}\\
(b) Department of Physics and Astronomy, University of British Columbia,\\ 
Vancouver, BC V6T~1Z1, Canada
%
%\end{flushleft}
%
\vspace{0.5cm}\\
email: nblinov@triumf.ca, dmorri@triumf.ca
\vspace{0.2cm}
\end{it}

\end{center}

\center{\today}
%\maketitle

\begin{abstract}

  In the Minimal Supersymmetric Standard Model~(MSSM), 
a Higgs boson mass of 125 GeV can be obtained with moderately
heavy scalar top superpartners provided they are highly mixed.
The source of this mixing, a soft trilinear stop-stop-Higgs coupling, 
can result in the appearance of charge- and color-breaking minima in 
the scalar potential of the theory.  If such a vacuum exists and
is energetically favorable, the Standard Model-like vacuum can decay 
to it via quantum tunnelling.  In this work we investigate the conditions
under which such exotic vacua arise, and we compute the tunnelling
rates to them.  Our results provide new constraints on the scalar top
quarks of the MSSM.

\end{abstract}

\end{titlepage}

\setcounter{page}{2}

\end{comment}
%%%%%%%%%%%%%%%%%%%%%%%%%%%%%%%%%%%%%%%%%%%%%%%%%%%%%%%%%%%%%%%%%%%%%%

\section{Introduction\label{sec:intro}}

   Supersymmetry predicts a scalar superpartner for every fermion
in the Standard Model~(SM)~\cite{Martin:1997ns}.  
While these scalar fields help to protect
the scale of electroweak symmetry breaking from large quantum corrections,
they can also come into conflict with existing experimental bounds.
This tension is greatest for the scalar top quarks~(stops).
On the one hand, the stops must be heavy enough to have avoided detection
in collider searches.  On the other hand, smaller stop masses maximize
the quantum protection of the electroweak 
scale~\cite{Barbieri:1987fn,Wymant:2012zp}. 

  In the minimal supersymmetric extension of the Standard Model~(MSSM),
there is an additional constraint on the stops implied by the discovery 
of a Higgs boson with mass near $m_h =125\,\gev$~\cite{higgs-atlas,higgs-cms}.
Specifically, the stops must be heavy enough to push the (SM-like) Higgs
mass up to the observed value~\cite{Hall:2011aa,Cao:2012fz}.
After electroweak symmetry breaking, the two gauge-eigenstate stops 
$\tilde{t}_L$ and $\tilde{t}_R$ mix to form two mass eigenstates, 
$\tilde{t}_1$ and $\tilde{t}_2$ ($m_{\tilde{t}_1}\leq m_{\tilde{t}_2}$).  
The corresponding mass-squared matrix in the $(\stp_L\;\stp_R)^T$ basis 
is~\cite{Martin:1997ns}
\beq
\mathcal{M}_{\tilde{t}}^2 = \left(
\begin{array}{cc}
m_{Q_3}^2+m_t^2 + D_L&m_tX_t\\
m_tX_t^*&m_{U_3}^2+m_t^2+D_R
\end{array}
\right) \ ,
\eeq
where $X_t = (A_t^* - \mu\cot\beta)$ is the stop mixing parameter,
$m_{Q_3,U_3}^2$ and $A_t$ are soft supersymmetry-breaking parameters, 
$\mu$ is the Higgsino mass parameter, 
$\tan\beta = v_u/v_d$ is the ratio of the two Higgs expectation values,
and $D_{L,R} = (t^3-Qs^2_W)m_Z^2\cos2\beta$ are the D-term contributions.
The stops generate the most important quantum corrections to the mass
of the SM-like Higgs state $h^0$ in the MSSM.
Decoupling the heavier Higgs bosons ($m_A \gg m_Z$), the $h^0$
mass at one-loop order is~\cite{Ellis:1990nz,Haber:1990aw,Carena:2000dp}
\beq
m_h^2 \simeq m_Z^2\cos^22\beta 
+ \frac{3}{4\pi^2}\frac{m_t^4}{v^2}\left[\ln\lrf{M_S^2}{m_t^2}
+\frac{X_t^2}{M_S^2}\left(1-\frac{X_t^2}{12M_S^2}\right)\right]
 \ ,
\label{eq:mh1loop}
\eeq
where $M_S = (m_{Q_3}m_{U_3})^{1/2}$. The first term is the tree-level 
contribution and is bounded above by $m_Z^2$.
The second term in Eq.~\eqref{eq:mh1loop} is the sum of one-loop top 
and stop contributions.  This correction is essential to raising
the mass of the SM-like MSSM Higgs mass to the observed value.

  The contribution of the stops to the $h^0$ mass depends on both the
mass eigenvalues and the mixing angle.  Without left-right stop mixing, 
at least one of the stops must be very heavy, $m_{\tilde{t}} \gtrsim 5\;\tev$, 
to obtain $m_h \simeq 125\,\gev$~\cite{Draper:2011aa}.  
This leads to a significant tension with the naturalness of the weak
scale~~\cite{Barbieri:1987fn,Wymant:2012zp}.  
This tension can be reduced by stop mixing, 
with the largest effect seen in the vicinity of the \emph{maximal mixing} 
scenario of $X_t \simeq \pm \sqrt{6}M_S$~\cite{Casas:1994us}. 
However, such large values of $X_t/M_S$ require a large value of 
$A_t$ (small $\mu$ is needed for naturalness~\cite{Kitano:2006gv})
which can induce new vacua in the scalar field space where
the stops develop vacuum expectation values (VEVs). 
The lifetime for tunnelling to these charge- and color-breaking (CCB) vacua 
must be longer than the age of the Universe to be consistent with our existence. 

  The existence of CCB stop vacua in the MSSM has been studied extensively
~\cite{Frere:1983ag,Gamberini:1989jw,Claudson:1983et,Casas:1995pd,Falk:1995cq,Riotto:1995am,Kusenko:1996jn,Casas:1996zy,LeMouel:2001sf,LeMouel:2001ym}. 
Under the assumption of $SU(3)_C\times SU(2)_L\times U(1)_Y$ $D$-flatness, 
an approximate analytic condition for the non-existence of a CCB stop vacuum 
is~\cite{Claudson:1983et,Casas:1995pd}
\beq
A_t^2 < 3(m_{Q_3}^2+m_{U_3}^2 + m_2^2) \ ,
\label{eq:analyticalbound}
\eeq
where $m_2^2 = m_{H_u}^2 + |\mu|^2$ and $m_{H_u}^2$ is the 
$H_u$ soft mass squared parameter.
Generalizations to less restrictive field 
configurations~\cite{Casas:1995pd,Casas:1996zy,LeMouel:2001sf,LeMouel:2001ym} 
and studies of the thermal evolution of such 
vacua~\cite{Carena:1996wj,Cline:1999wi,Patel:2013zla} have been 
performed as well. 
Relaxing the requirement of absolute stability of our electroweak
vacuum and demanding only that the tunnelling rate to the CCB vacua is 
sufficiently slow provides a weaker bound.  The tunnelling rate was computed
in Ref.~\cite{Kusenko:1996jn}, where the net requirement for metastability
was expressed in terms of the empirical relation
\beq
A_t^2  +3\mu^2 \lesssim 7.5(m_{Q_3}^2+m_{U_3}^2) \ .
\label{eq:empiricalbound}
\eeq
In this work we attempt to update and clarify the stability 
and metastability bounds on the parameters in the stop sector of the MSSM.
We expand upon the previous body of work by investigating the 
detailed dependence of the limits on the underlying set of stop parameters.
Furthermore, we relate our revised limits to recent Higgs and stop
search results at the LHC.

  The outline of this paper is as follows.  In Section~\ref{sec:potential}
we specify the ranges of MSSM parameters and field configurations to 
be considered.  Next, in Section~\ref{sec:stability} we investigate 
the necessary conditions on the underlying stop and Higgs parameters 
for the scalar potential to be stable or safely metastable.  
We then compare the constraints from metastability to existing limits
on the MSSM stop parameters from the Higgs mass in Section~\ref{sec:hmass},
as well as to direct and indirect stop searches in Section~\ref{sec:bounds}.
Finally, we conclude in Section~\ref{sec:conc}.
Some technical details related to our tunnelling calculation are 
expanded upon in the Appendix.

\section{Parameters and Potentials~\label{sec:potential}}

  In our study, we consider only variations in the scalar
fields derived from the superfields 
$Q_3 \to (\stp_L,\tilde{b}_L)^T$, $U^c_3\to \stp_R^*$, $H_u\to(H_u^+,H_u^0)$,
and $H_d \to (H_d^0,H_d^-)$.  To make this multi-dimensional space
more tractable, we further restrict ourselves to configurations
where $\tilde{b}_L = H_u^+ = H_d^- = 0$ and the remaining fields
(and MSSM parameters) are real-valued. 
Previous studies of CCB vacua in the stop direction suggest that 
this condition is not overly restrictive~\cite{Casas:1995pd}.

\subsection{Scalar Potential}

  Under these assumptions, the tree-level scalar potential becomes
\beq
V_{\mathrm{tree}} = V_2 + V_3 + V_4
\label{eq:vtree}
\eeq
where
\beq
V_2 &=& (m_{H_u}^2+|\mu|^2)(H_u^0)^2 + (m_{H_d}^2+|\mu|^2)(H_d^0)^2
- 2bH_u^0H_d^0 + m_{Q_3}^2\stp_L^2 + m_{U_3}^2\stp_R^2 \\
V_3 &=& 2y_t(A_tH_u^0 -\mu\,H_d^0)\,\stp_L\stp_R\\
V_4 &=& y_t^2\left[\stp_L^2\stp_R^2+\stp_L^2(H_u^0)^2+\stp_R^2(H_u^0)^2\right]
+ V_D \ ,
\eeq
with 
\beq
V_D = \frac{{g'}^2}{8}\left[(H_u^0)^2-(H_d^0)^2+\frac{1}{3}\stp_L^2
-\frac{4}{3}\stp_R^2\right]^2 
+ \frac{g^2}{8}\left[-(H_u^0)^2+(H_d^0)^2+\stp_L^2\right]^2
+ \frac{g_3^2}{6}\left(\stp_L^2-\stp_R^2\right)^2 \ .
\eeq
In writing this form, we have implicitly assumed that the stops
are aligned (or anti-aligned) in $SU(3)_C$ space, so that $\stp_L$ and $\stp_R$
may be regarded as the magnitudes of these color vectors (up to a possible
sign).  It is not hard to show that such an alignment maximizes the likelihood
of forming a CCB minimum.

  In our analysis of metastability, we use the tree-level potential 
of Eq.~\eqref{eq:vtree} with the parameters in it taken to be 
their $\drbar$ running values defined at the scale $M_S$.  
However, we also compare our metastability results to a full 
two-loop calculation of the Higgs boson mass.  While this is a mismatch 
of orders, we do not expect that including higher order corrections 
will drastically change our metastability results for two reasons.  
First and most importantly, the formation of CCB vacua is driven 
by the trilinear stop coupling $A_t$, 
which is already present in the tree-level potential.  
Second, when a CCB vacuum exists, the large stop Yukawa coupling $y_t \sim 1$ 
implies that it typically occurs at field values on the order 
of $M_S$~\cite{Casas:1995pd}.  Thus we do not expect large logarithmic 
corrections from higher orders.  

  Including higher-order corrections in the tunnelling analysis is
also challenging for a number of technical reasons.  
Turning on multiple scalar fields, the mass matrices entering 
the Coleman-Weinberg corrections to the effective potential become 
very complicated and multi-dimensional~\cite{Cline:1999wi}.  
These corrections can be absorbed into running couplings 
by an appropriate field-dependent choice of the renormalization 
scale~\cite{Gamberini:1989jw}.  In doing so, however, the otherwise 
field-independent corrections to the vacuum energy 
(which are not included in the Coleman-Weinberg potential)
develop a field dependence.  These vacuum energy corrections 
must be included to ensure the net scale independence of the effective 
potential~\cite{Kastening:1991gv,Ford:1992mv}.  
Beyond the effective potential, kinetic corrections
(\emph{i.e.} derivative terms in the effective action)
will also be relevant for the non-static tunnelling configurations
to be studied.  Furthermore, the effective potential and the
kinetic corrections are both gauge dependent~\cite{Jackiw:1974cv,Patel:2011th}.
The gauge dependence of the effective potential can be shown to 
cancel on its own for static points~\cite{Nielsen:1975fs,Fukuda:1975di}.  
However, to ensure the gauge invariance
of the non-static tunnelling configuration and thus the decay rate, 
kinetic corrections must be included as well~\cite{Metaxas:1995ab,Garny:2012cg}.
For these various reasons, we defer an investigation of higher-order
corrections to metastability to a future work.

\subsection{Parameter Ranges\label{sec:para}}

  Without loss of generality, we may redefine $H_u^0$ and $H_d^0$ such
that $b$ and $H_u^0$ are both positive.  This ensures that the unique 
SM-like vacuum (with $\stp_L = \stp_R = 0)$ has 
$\tan\beta = \langle H_u^0\rangle/\langle H_d^0\rangle >0$,
and thus $\langle H_d^0\rangle > 0$ as well. 
By demanding that a local SM-like vacuum exists, $b$, $m_{H_u}^2$, 
and $m_{H_d}^2$ can be exchanged in favour of 
$v = \sqrt{\langle H_u^0\rangle^2+\langle H_u^0\rangle^2}$, $\tan\beta$,
and the pseudoscalar mass $m_A$:
\beq
b &=& \frac{1}{2}m_A^2\sin({2\beta})\\
m_{H_u}^2 &=& -\mu^2 + m_A^2\cos^2\!\beta + \frac{1}{2}m_Z^2\cos(2\beta)\\
m_{H_d}^2 &=& -\mu^2 + m_A^2\sin^2\!\beta - \frac{1}{2}m_Z^2\cos(2\beta) \ .
\eeq
Moving out in the stop directions, we may also redefine $\stp_L$ and 
$\stp_R$ such that $\stp_L$ is positive. 

  The parameter ranges we investigate are motivated by existing bounds
on the MSSM and naturalness.  We typically scan over 
$(m_{Q_3}^2,X_t)$ while holding other potential parameters fixed.
We also consider discrete variations in $m_{U_3}^2/m_{Q_3}^2$,
$\tan\beta$, $\mu$, and $m_A$.  The corresponding ranges are
specified in Table~\ref{tab:scanrange}.   
For the remaining supersymmetry breaking parameters, we choose
$m_{\tilde{f}} = 2\,\tev$ and $A_f = 0$ for all sfermions other than the stops, 
as well as $M_1 = 300\,\gev$, $M_2 = 600\,\gev$, and $M_3 = 2\,\tev$.  
To interface with the Higgs mass calculation, we take these to be running
$\drbar$ values defined at the input scale $M_S = (m_{Q_3}m_{U_3})^{1/2}$.
We also use running $\drbar$ values of $y_t$, $g'$, $g$, and $g_3$
at scale $M_S$ when evaluating the potential.

\begin{table}
\centering
    \begin{tabular}{| c | c |}
    \hline
    Parameter & Values \\
    \hline
    $|m_{Q_3}|$ & $[300,3000]\;\gev$ \\ 
    $m_{Q_3}^2/m_{U_3}^2$ & ${0.3\,,1,\,3}$\\ 
    $X_t$ & $[-10,10]\times |m_{Q_3}|$\\
    $\mu$ & ${150,\,300,\,500}\;\gev$\\
    $m_A$ & ${1000}\;\gev$\\
    $\tan\beta$ & ${5,\,10,\,30}$\\
    \hline
  \end{tabular}
\caption{MSSM scalar potential parameter scan ranges. 
The values of other parameters to be considered
are described in the text. \label{tab:scanrange}}.
\end{table}

\section{Limits From Vacuum Stability~\label{sec:stability}}

  A necessary condition on the viability of any realization of the MSSM
is that the lifetime of the SM-like electroweak vacuum at zero temperature
be longer than the age of the Universe.  This will certainly be the case 
if the electroweak vacuum is a global minimum, and it can also be true 
in the presence of a deeper CCB minimum provided the tunnelling rate 
is sufficiently small.  More stringent conditions can be derived 
for specific cosmological histories~\cite{Cline:1999wi}.  While color-broken
phases in the early Universe can have interesting cosmological implications,
such as for baryogenesis~\cite{Carena:1996wj,Cline:1999wi,Patel:2013zla},
we focus exclusively on the history-independent $T= 0$ 
metastability condition.

\subsection{Existence of a CCB Vacuum}

  The first step in a metastability analysis is to determine whether a CCB
minimum exists.  Such minima are induced by a competition between the 
trilinear $A$ and quartic couplings $\lambda$  in the potential, and one generally expects
$\langle \phi \rangle_{\mathrm{CCB}}\sim A/\lambda$~\cite{Casas:1995pd}.
We use this expectation as a starting point for a numerical minimization 
of the potential, Eq.~\eqref{eq:vtree}, employing the minimization 
routine Minuit2~\cite{James:1975dr}. For every MSSM model, 
we choose the starting point to be 
$\langle \phi_i \rangle_{\mathrm{CCB}} = \xi_i A_t$, 
where $\xi_i\in[-1,1]$ is chosen randomly. 
The global CCB vacua we find are generally unique, up to our restrictions
of $H_u^0,\,\tilde{t}_L\geq 0$. 
If no global CCB minimum is found, the minimization is repeated several 
times with new $\xi_i$ values. If the global minimum turns out 
to be the EW vacuum, the model is considered to be Standard Model-like~(SML).

\subsection{Computing the Tunnelling Rate}

  When a deeper CCB vacuum is found, the decay rate of the SML vacuum
is computed using the Callan-Coleman 
formalism~\cite{Coleman:1977py, Callan:1977pt}, 
where the path integral is evaluated in the semi-classical approximation. 
The decay rate per unit volume is given by 
\begin{equation}
\Gamma/V = C \exp(-B/\hbar) \ ,
\label{eq:decayrate}
\end{equation}
where $B=S_E[\bvec\phi]$ is the Euclidean action evaluated
on the bounce solution $\bvec{\phi}$.  The bounce is $O(4)$-symmetric,
depending only on $\rho = \sqrt{t^2+\vec{x}^2}$, and satisfies the classical
equations of motion subject to the boundary conditions
$\del_{\rho}\bvec{\phi}|_{\rho=0} = 0$ and 
$\lim_{\rho\to\infty}\bvec{\phi} = \vec{\phi}_+$, where $\vec{\phi}_+$ is the
false-vacuum field configuration.
The pre-exponential factor $C$ is obtained from fluctuations around
the classical bounce solution.  It is notoriously difficult to 
compute~\cite{Dunne:2005rt,Min:2006jg},
and is therefore usually estimated on dimensional grounds~\cite{Weinberg:1992ds}.
We use
\beq
[C] = M^4\;\Rightarrow C = (100\;\gev)^4 \ .
\eeq
The metastability of the SM-like vacuum then requires 
\begin{equation}
\Gamma^{-1} \gtrsim t_0\;\Rightarrow\;B/\hbar \gtrsim 400,
\label{eq:actionconstraint}
\end{equation}
where $t_0 = 13.8\;\mathrm{Gyr}$ is the age of the universe.
Our choice of scale for $C$ corresponds to the SM-like vacuum,
and provides a reasonable lower bound on $C$.  Larger values of
$C$ would increase the decay rate, implying that the limits
we derive are conservative.

  Finding the bounce $\bvec\phi$ is straightforward in one field dimension,
since the equation of motion can be solved by the shooting method.
This method reduces the problem to a root-finding task for the 
correct boundary conditions and relies on the unique topology of the
one-dimensional field space.  Unfortunately, this strategy becomes 
intractable with more than one field dimension.  Several methods of solving 
the multi-field bounce equation of motion have been 
proposed~\cite{Kusenko:1995jv,Cline:1999wi,Konstandin:2006nd,Park:2010rh}.
In the present analysis we use the public code 
CosmoTransitions~\cite{Wainwright:2011kj}.\footnote{We modify the 
code slightly, replacing an instance of \texttt{scipy.optimize.fmin} 
by \texttt{scipy.optimize.fminbound} in the class 
\texttt{pathDeformation.fullTunneling}. This allows CosmoTransitions 
to better deal with very shallow vacua. The same modification has 
been used in Ref.~\cite{Reece:2012gi} (see Footnote 1).}

  CosmoTransitions~(CT) implements a path deformation method similar
to the that suggested in Ref.~\cite{Cline:1999wi}.  Once a pair of local 
minima are specified, CT fixes a one-dimensional path between them 
in the field space.
Along this path, the one-dimensional bounce solution can be computed 
using the shooting method.  In Appendix~\ref{sec:stuff}, we show that 
the action computed from the bounce solution for any such fixed path 
is necessarily greater than or equal to the unconstrained
bounce action.  The fixed path in field space is then deformed by
minimizing a set of perpendicular gradient terms to be closer to the true
bounce path through the field space.  This procedure is iterated until 
convergence is reached.  We exclude any points where CT fails to converge.

  This path deformation approach has several advantages over other methods. 
Here, the bounce equation of motion is solved directly, while many 
other approaches involve minimization of a discretized action 
as part of the procedure.  This is numerically costly, since one needs 
both a fine lattice spacing to evaluate derivatives accurately, 
and a large $\rho$ domain to accommodate the boundary condition at infinity.  
Path deformation involves no discretization or large-scale minimization.  
As a result CosmoTransitions is quite fast for our four-field 
tunnelling problem. 

  We also cross check the CT results in two ways.  First, we have
 compared CT to the discretized action methods of
Refs.~\cite{Konstandin:2006nd,Park:2010rh} for a set of special
cases, and we generally find agreement between these approaches.  
Second, we also compute the bounce action independently along the 
optimal path determined by CT, allowing us to estimate the numerical 
uncertainty on the bounce.  Finally, let us emphasize once more that 
even if the path determined by CT is not the true
tunnelling trajectory, our result in Appendix~\ref{sec:stuff} 
implies that it still provides an upper bound on the bounce action,
and thus a lower bound on the tunnelling rate.  

We note that recently a new program, \verb=Vevacious=~\cite{Camargo-Molina:2013qva}, 
has been released that can also be used to study metastability in field theories 
with many scalar fields. While we do not use this code, we share some similarities 
with their approach in that we both employ Minuit for potential minimization and 
CT for tunneling rates. However, as mentioned above, we also carried out 
extensive independent checks of the tunneling calculation.

\subsection{Results and Comparison\label{sec:results}}

\begin{figure}[ttt]
\centering
\if\withFigures1
\includegraphics[width=0.5\textwidth]{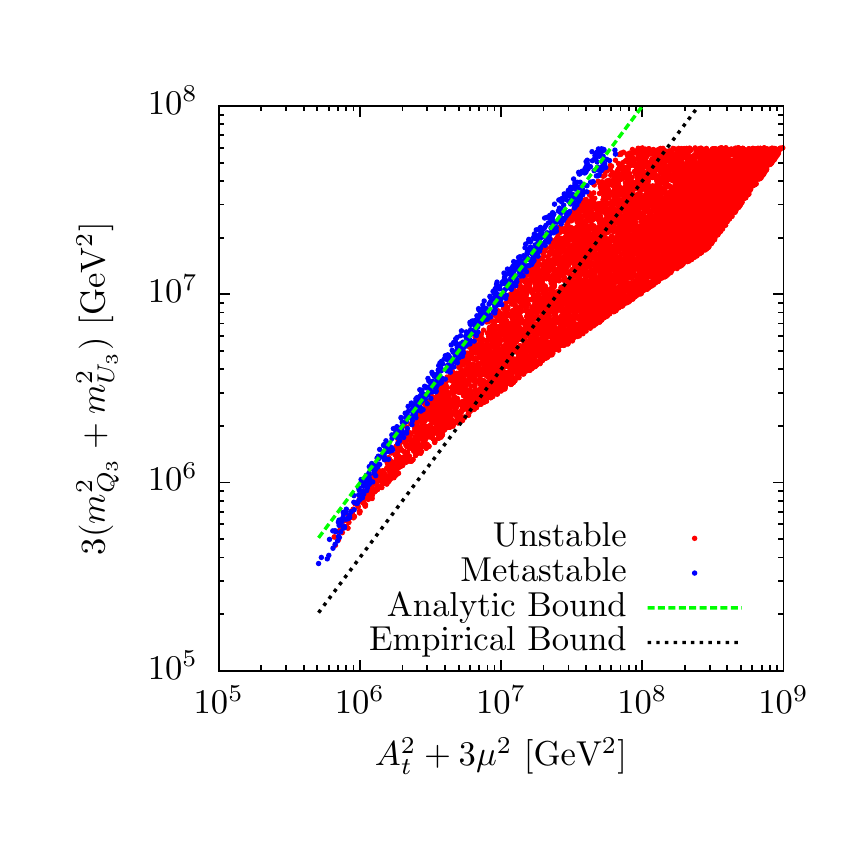}
\fi
\caption{Limits from metastability and the existence of a local
SM-like~(SML) vacuum alone for $\tan\beta = 10$, $\mu = 300\,\gev,$
$m_A = 1000\,\gev$, and $m_{U_3}^2=m_{Q_3}^2$.  
All points shown have a global CCB minimum and a 
local SML minimum.  The red points are dangerously unstable, 
while the blue points are consistent with metastability.
The green dashed line is the analytic bound of Eq.~\eqref{eq:analyticalbound} 
and the black dotted line corresponds to Eq.~\eqref{eq:empiricalbound},
the empiricial bound from Ref.~\cite{Kusenko:1996jn}. The values of
the other MSSM parameters used here are described in the text. 
\label{fig:nomh_default}}
\end{figure}

  We begin by presenting our limits from metastability alone, without
imposing any other constraints such as the Higgs mass requirement.
This allows for a direct comparison with the results of 
Ref.~\cite{Kusenko:1996jn}.  In Fig.~\ref{fig:nomh_default} 
we show a scan over $X_t$ and $m_{Q_3}^2$ while keeping fixed $m_A = 1000\;\gev$, 
$\tan\beta=10$, $\mu = 300\;\gev$, and $m_{U_3}^2/m_{Q_3}^2 = 1$. 
Every point shown is a model with a global CCB vacuum.
The red points have a tunnelling action $B/\hbar < 400$, 
and are therefore unstable on cosmological time scales.
The blue points have a metastable SM-like vacuum with $B/\hbar > 400$. 
Also shown in the figure is the analytic bound 
(green dashed line) of Eq.~\eqref{eq:analyticalbound}, 
and the empirical result (black dotted line) from Ref.~\cite{Kusenko:1996jn}
given in Eq.~\eqref{eq:empiricalbound}. 

  The shape of the regions shown in Fig.~\ref{fig:nomh_default} can be
understood simply.  As expected, the existence of a CCB vacuum requires
a large value of $A_t/M_S$.  The cutoff at the upper-left diagonal 
edge corresponds to the absence of a CCB vacuum.  Above and to the left of 
this boundary, the SML minimum is a global one and the EW vacuum 
can be absolutely stable.  There is also a lack of points below a
lower-right diagonal edge.  Here, one of the physical stops becomes tachyonic,
and the SML vacuum disappears altogether. At low values of 
$A_t^2$, we see that the CCB region is squeezed between 
the SML region (on the left) and the tachyonic stop region (on the right), 
giving rise to the cutoff seen in the lower left corner.

  It is apparent from Fig.~\ref{fig:nomh_default} that we find much
more restrictive metastability bounds on the MSSM than the empirical
relation of Eq.~\eqref{eq:empiricalbound} from Ref.~\cite{Kusenko:1996jn}.
We also see that the analytic bound of Eq.~\eqref{eq:analyticalbound}
tends to underestimate the existence of CCB vacua, and that it accidentally
lines up fairly well with the lower boundary of metastability.
It is not clear why our results should be so much more restrictive
than those found in Ref.~\cite{Kusenko:1996jn}, but we are confident 
that the path deformation method of CT 
(and our several cross-checks) gives a robust upper bound on the bounce action.  
We find qualitatively similar results for the other parameters ranges
described in Table~\ref{tab:scanrange}.  The quantitative results for
these ranges will be presented in more detail below in the context of the
Higgs mass.

\section{Implications for the MSSM Higgs Boson\label{sec:hmass}}

  As discussed in the Introduction, there is a significant tension
in the MSSM between obtaining the observed Higgs boson mass and keeping
the stops relatively light.  This tension is reduced when the stops
are strongly mixed.  To obtain such mixing, large values of $X_t$ are needed.
We have just seen that large values of $X_t$ can lead to dangerous
CCB minima.  In this section we compare the relative conditions imposed
by each of these requirements.

  To calculate the physical $h^0$ Higgs boson mass, we use 
FeynHiggs~2.9.5~\cite{Heinemeyer:1998yj}. 
We also use this program together with SuSpect~2.43~\cite{Djouadi:2002ze}
to compute the mass spectrum of the MSSM superpartners.
As inputs, we take $m_t^{\mathrm{pole}} = 173.1\,\gev$ 
and $\alpha_s(m_Z)=0.118$~\cite{Beringer:1900zz}.
Our results are exhibited in terms of variations on the fiducial
MSSM parameters $\tan\beta = 10$, $\mu = 300\,\gev,$
$m_A = 1000\,\gev$, and $m_{U_3}^2=m_{Q_3}^2$.
The other MSSM parameters are taken as in Section~\ref{sec:para}.

\begin{figure}
\centering
\if\withFigures1
\includegraphics[width=0.5\textwidth]{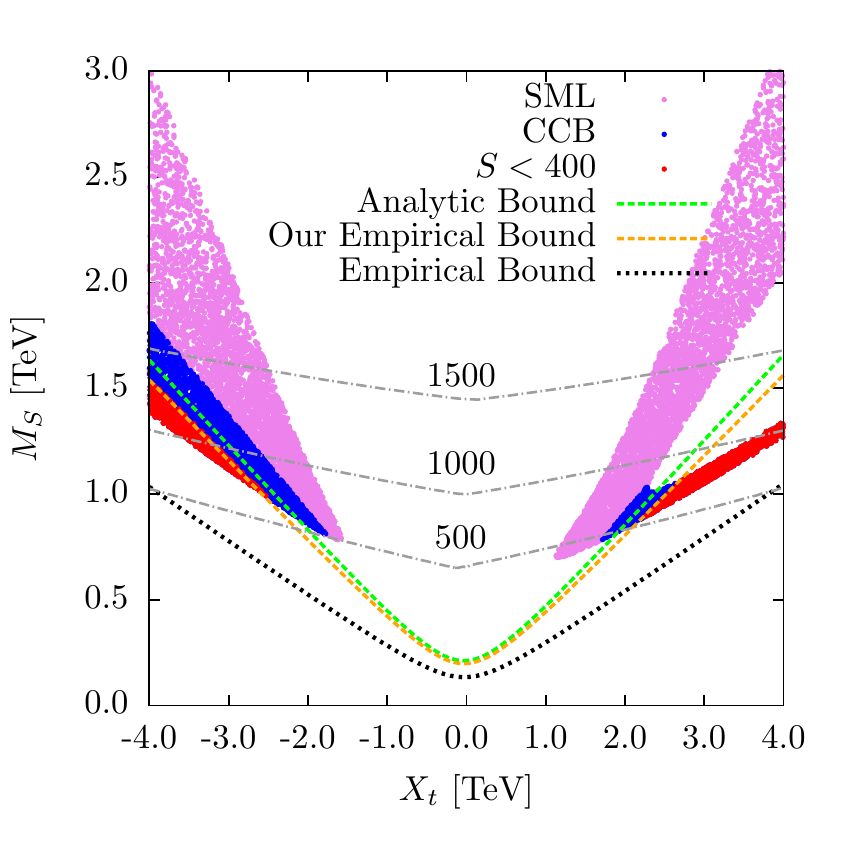}
\fi
\caption{Metastability bounds relative to the MSSM Higgs mass.
The colored bands contain models for which $123\;\gev< m_h < 127\;\gev$.  
Pink models have an absolutely stable SML vacuum, 
blue points have a global CCB minimum, while red points are unstable 
on cosmological time scales. The green 
dashed line is the analytic bound of Eq.~\eqref{eq:analyticalbound} 
and the black dotted line is Eq.~\eqref{eq:empiricalbound}. The 
orange dashed line is an approximate empirical bound discussed in 
Appendix~\ref{sec:empbound}.
The grey 
dot-dashed contours are lines of constant lightest stop mass (in GeV). 
MSSM parameters used here are described in the text. 
\label{fig:mh_default}}
\end{figure}

  In Fig.~\ref{fig:mh_default} we show points in the  $X_t$-$M_S$ 
plane (where $M_S = (m_{Q_3}m_{U_3})^{1/2}$) that produce a Higgs mass 
in the range $123\;\gev< m_h < 127\;\gev$.  All other parameters are 
set to their fiducial values described above.
The pink (blue) region are models with a global SML (CCB) vacuum.
The red points are excluded by metastability.  The dashed lines
show the approximate CCB condition of Eq.~\eqref{eq:analyticalbound}, 
the empirical limit of Eq.~\eqref{eq:empiricalbound}, and our own
attempt at an empirical limit on metastability to be discussed below.
The requirement of metastability cuts off a significant portion of
the allowed range at very large $|X_t|$. Also shown are contours of 
constant $m_{\stp_1}$, the lightest stop mass (grey dot-dashed lines).

\begin{figure}[!ttt]
\centering
\if\withFigures1
\includegraphics[width=0.4\textwidth]{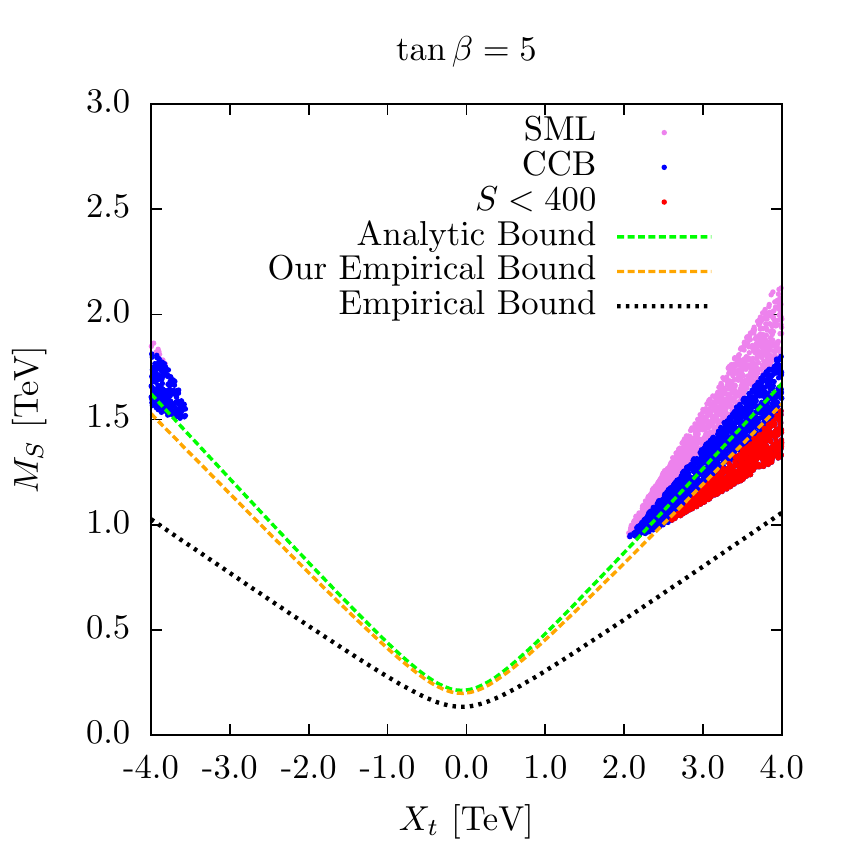}
\includegraphics[width=0.4\textwidth]{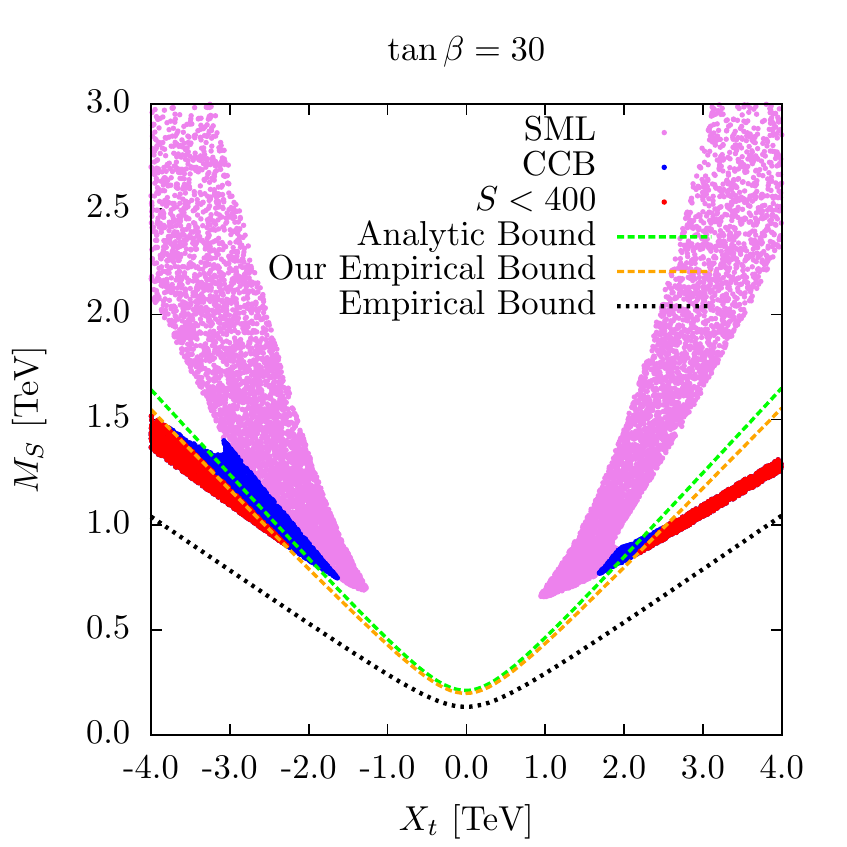}\\
\includegraphics[width=0.4\textwidth]{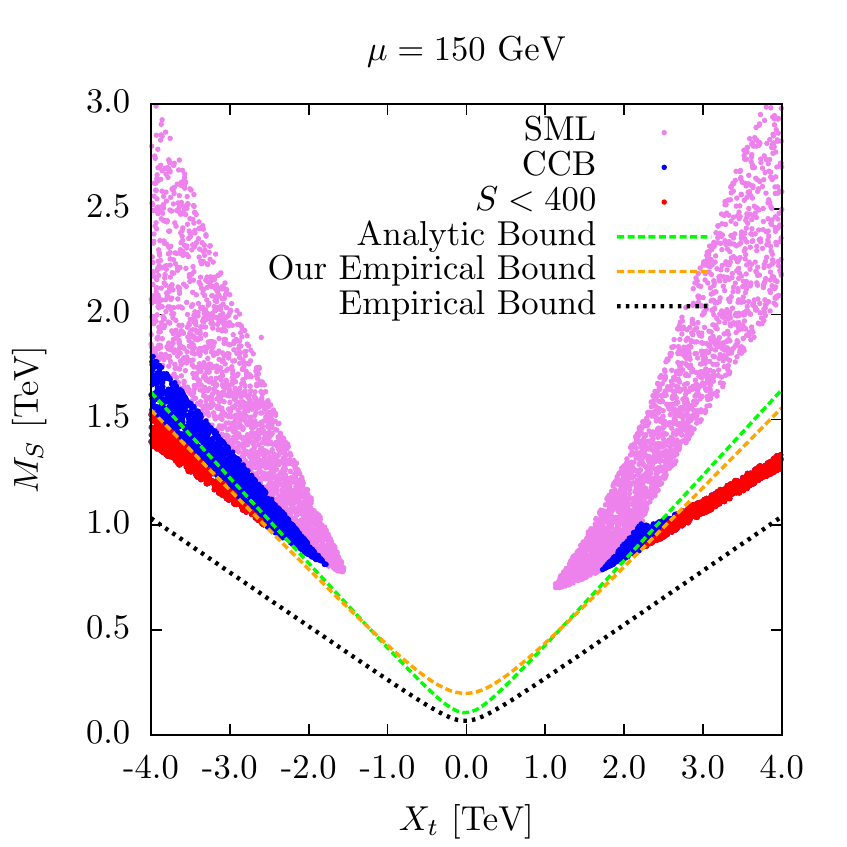}
\includegraphics[width=0.4\textwidth]{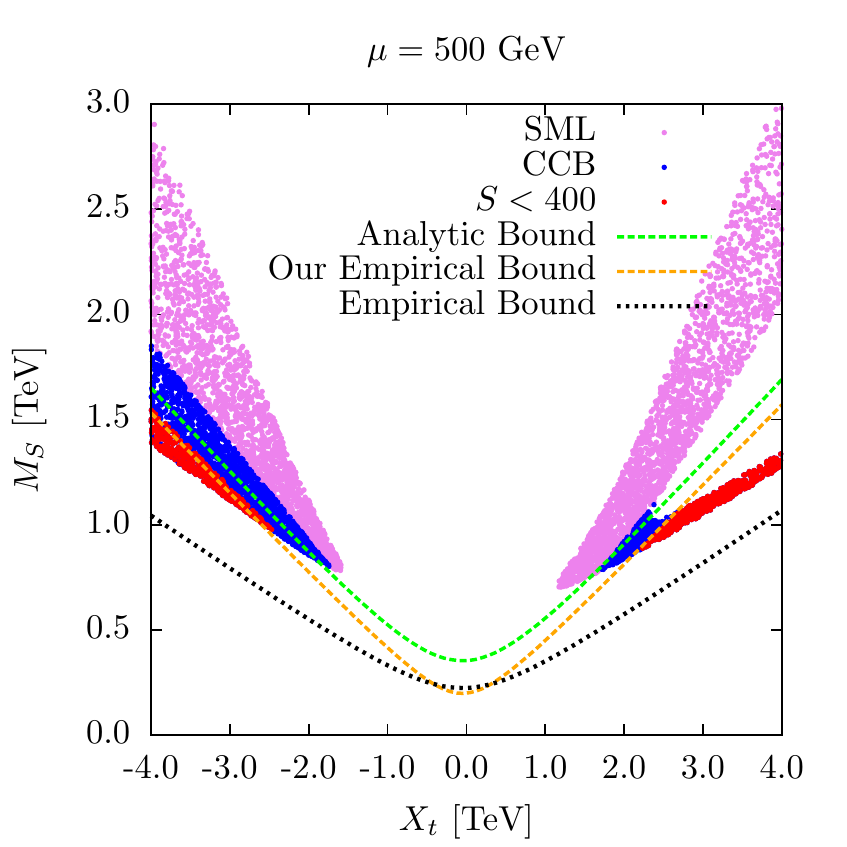}\\
\includegraphics[width=0.4\textwidth]{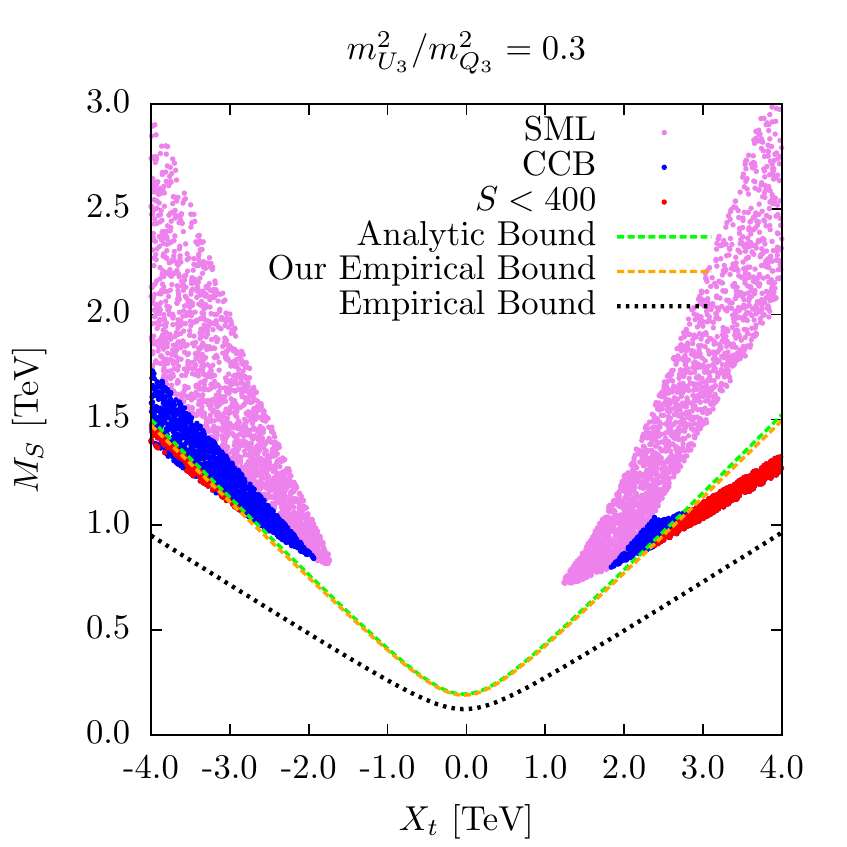}
\includegraphics[width=0.4\textwidth]{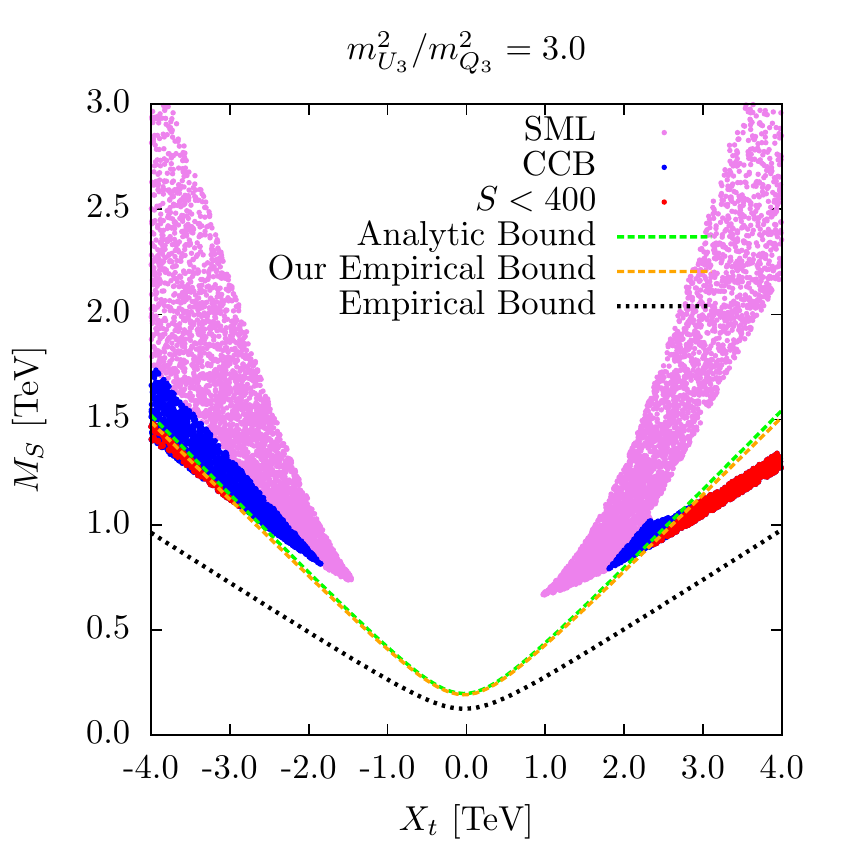}
\fi
\caption{Metastability with the correct Higgs mass, $123<m_h<127\,\gev$.
The labelling is the same as in Fig.~\ref{fig:mh_default}, and the
relevant MSSM parameter parameters are varied one at a time as
summarized in Table~\ref{tab:scanrange}.
\label{fig:mh_variations}}
\end{figure}

  In Fig.~\ref{fig:mh_variations} we show the additional dependence of 
the Higgs mass and the metastability bounds on other relevant
MSSM parameters.  All parameters are set to their fiducial values
except for those we vary one at a time.
In the top row we show results for $\tan\beta = 5~(30)$ 
on the left (right).  Reducing $\tan\beta$ decreases the tree-level
contribution to the MSSM Higgs mass, and so larger values of $M_S$ are
needed to raise $m_h$ to the observed range.  These larger values also
lead to shallower CCB minima and lower tunnelling rates.  Larger values
of $\tan\beta$ do not appear to differ much from $\tan\beta =10$.

  In the middle row of Fig.~\ref{fig:mh_variations} we show results for
$\mu = 150~(500) \ ,\gev$ on the left (right).  We do not see a large
amount of variation in the exclusions from metastability, which is not
surprising given that generally have $X_t \simeq A_t \gg \mu$.  Setting
$\mu = -300\,\gev$ also produces very similar results.

  In the bottom row of Fig.~\ref{fig:mh_variations} we show the 
same metastability limits for $m_{U_3}^2/m_{Q_3}^2 = 0.3~(3.0)$ 
on the left (right).  For these unequal values, there is a 
tension between minimizing the quadratic terms in the potential 
and reducing the quartic terms through $SU(3)_c$ $D$-flatness.
Unequal squark VEVs also tend to reduce the effective trilinear term.
Together, these effects reduce the metastability constraint somewhat,
but do not eliminate it.  

  In summary, the constraint imposed by CCB metastability rules out 
a significant portion of the MSSM stop parameter space that can produce 
a Higgs mass near the observed value.  The limits are strongest 
on the outer branches at large $|X_t|$.  Varying other MSSM parameters
within the restricted ranges we have considered does not drastically 
alter this result.  By comparison, the empirical bound from 
Ref.~\cite{Kusenko:1996jn}
does not rule out any of the stop parameter space consistent with
the Higgs mass.  

  As a synthesis of these results, we have attempted to obtain an
improved empirical bound on stop-induced metastability.  We find
the approximate limit
\beq
A_t^2 \lesssim \left(
3.4 + 0.5\frac{|1-r|}{1+r}
\right)m_T^2 + 60\,m_2^2 
\label{eq:empus} \ ,
\eeq
where $m_T^2 = (m_{Q_3}^2+m_{U_3}^2)$, $m_2^2 = (m_{H_u}^2+\mu^2)$,
and $r = m_{U_3}^2/m_{Q_3}^2$.  Let us emphasize that this limit is
very approximate and only applies to smaller values of $\mu$,
larger values of $m_A$, moderate $\tan\beta$, and $r$ not too different 
from unity.  Details on the derivation of this bound are 
given in Appendix~\ref{sec:empbound}.

\section{Comparison to Other Stop Constraints\label{sec:bounds}}  

  The metastability conditions we find exclude parameter
regions with large stop mixing.  This mixing can produce one relatively 
light stop mass eigenstate as well as a significant mass splitting 
between the members of the $\widetilde{Q}_3$ sfermion doublet.  
These features are constrained indirectly by electroweak and 
flavor measurements, as well as by direct searches for a light stop.  
In this section we compare these additional limits to the bounds from
metastability.

\subsection{Precision Electroweak and Flavor}

  The most important electroweak constraint on light stops comes from
$\Delta\rho$, corresponding to the shift in the $W$ mass relative
to the $Z$. In the context of highly mixed stops motivated by the Higgs mass,
this effect has been studied in Refs.~\cite{Barger:2012hr,Espinosa:2012in}. 
We have computed the shift $\Delta\rho$ due to stops and sbottoms
using  SuSpect~2.43~\cite{Djouadi:2002ze}, which applies the one-loop
results contained in Refs.~\cite{Drees:1990dx,Heinemeyer:2004gx}.
With a Higgs mass of $m_h \simeq 125\,\gev$, the preferred range is 
$\Delta\rho = (4.2\pm 2.7)\times 10^{-4}$~~\cite{Barger:2012hr}.

  Supersymmetry can also contribute to flavor-mixing.  
Assuming only super-CKM squark mixing 
(or even minimal flavor violation~\cite{D'Ambrosio:2002ex}), 
the most constraining flavor observable is frequently the branching 
ratio $\mathrm{BR}(B \to X_s\gamma)$.  It receives contributions
in the MSSM from stop-chargino and top-$H^+$ loops.  
These contributions tend to cancel each other such that 
the cancellation would be exact
in the supersymmetric limit~\cite{Ciuchini:1998xy}.  
With supersymmetry breaking, the result depends on the stop masses
and mixings, $\tan\beta$, $\mu$, and the pseudoscalar mass $m_A$.  
Constraints on light stops from $\mathrm{BR}(B\to X_s\gamma)$ were considered 
recently in Refs.~\cite{Heinemeyer:2011aa,Espinosa:2012in}.
The SM prediction is $\text{BR}(B\to X_s\gamma) 
= (3.15\pm 0.23)\times 10^{-4}$~\cite{Grzadkowski:2008mf},
while a recent Heavy Flavor Averaging Group compilation of 
experimental results finds $\text{BR}(B\to X_s\gamma) 
= (3.55\pm 0.24\pm 0.09)\times 10^{-4}$~\cite{Amhis:2012bh}.
We have investigated the limit from $\mathrm{BR}(B\to X_s\gamma)$ and 
other flavor observables using SuperIso~3.3~\cite{Arbey:2009gu}
assuming only super-CKM flavor mixing.
 
\begin{figure}[!ttt]
 \begin{center}
    \if\withFigures1
        \includegraphics[width=0.5\textwidth]{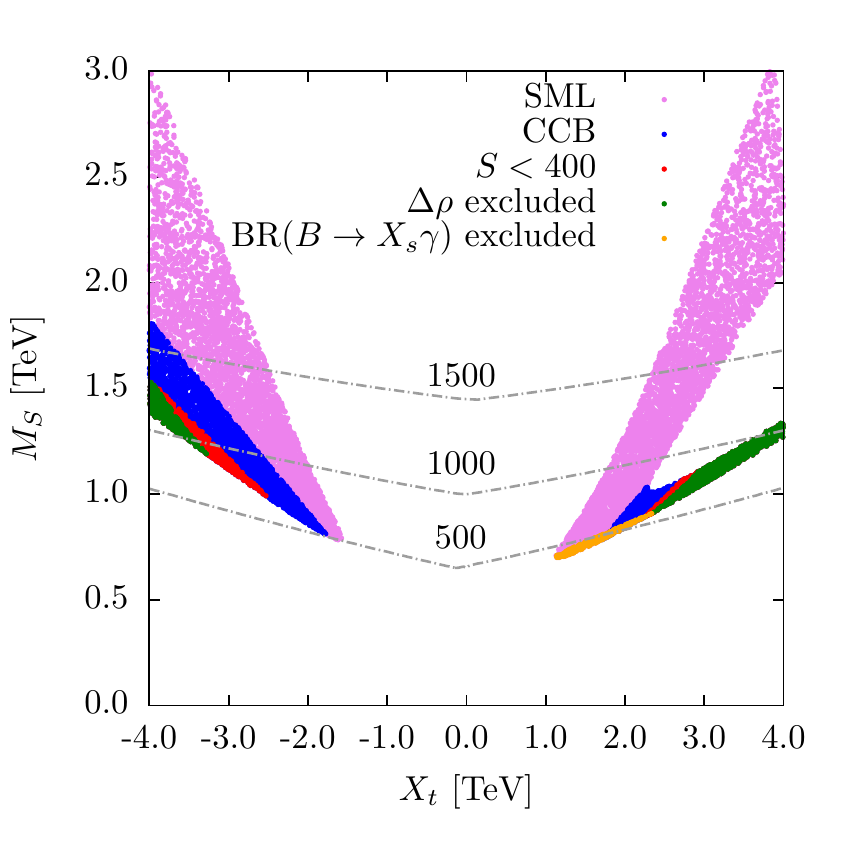}
     \fi
 \end{center}
 \caption{Points in the $X_t\!-\!m_{Q_3}$ plane with 
 $123\;\gev< m_h < 127\;\gev$ as well as exclusions from metastability (red points)
from precision electroweak $\Delta\rho$ (green points) and flavor $\mathrm{BR}(B\to X_s\gamma)$ (orange points).
The MSSM parameters used are the same as in Fig.~\ref{fig:mh_default}.
}
 \label{fig:flpew}
 \end{figure}

  In Fig.~\ref{fig:flpew}, we show the exclusions from flavor and electroweak
bounds for model points with $123\;\gev< m_h < 127\;\gev$ for $\tan\beta = 10$, 
and $m_A=1000\,\gev$, $\mu = 300\,\gev$, and $m_{Q_3}^2=m_{U_3}^2$
in the $X_t\!-\!m_{Q_3}$ plane. We impose the generous $2\sigma$ 
constraints $\Delta\rho\in[-1.2,9.4]\times 10^{-4}$ and
$\mathrm{BR}(B\to X_s\gamma)\in [2.86,4.24]\times 10^{-4}$ 
and show them together with the metastability constraint from 
the previous Section. The green points show the regions excluded by 
$\Delta\rho$ while the orange points show those excluded by
$\mathrm{BR}(B\to X_s\gamma)$.  

  The exclusion due to $\Delta\rho$ can be understood in terms of
the large stop mixing induced by $X_t$, which generates a significant 
splitting between the mass eigenstates derived from the 
$\widetilde{Q}_3=(\tilde{t}_L,\tilde{b}_L)^T$ $SU(2)_L$ doublet.
This constraint depends primarily on the stop parameters,
and is mostly insensitive to variations in $\mu$, $m_A$, and $\tan\beta$.
While this bound overlaps significantly with the limit from metastability,
there are regions where only one of the two constraints applies.
The limits from $\Delta\rho$ are also weaker for $m_{Q_3}^2 > m_{U_3}^2$.

  Limits from $\text{BR}(B\to X_s\gamma)$ are less significant
for this set of fiducial parameters with a moderate value of $\tan\beta$.
However, this branching fraction depends significantly on $\mu$, $m_A$,
and $\tan\beta$, and the limit can be much stronger or much weaker depending on
the specific values of these parameters.  We do not attempt to delineate
the acceptable parameter regions, but we do note that the constraint from
metastability can rule out an independent region of the parameter space.

\subsection{Direct Stop Searches}

  Stops have been searched for at the LHC in a diverse range of 
final states, and these studies rule out stop masses up to
200-600\!\gev, depending on how the stop decays (see 
\emph{e.g.} Refs.~\cite{Aad:2013ija,ATLAS:2013ama,Chatrchyan:2013xna,CMS:stp}). 
While the large stop mixing that occurs in the region excluded
by metastability considerations can produce lighter stops,
the stop masses in this dangerously metastable region are not 
necessarily light, as can be seen in Fig.~\ref{fig:mh_default}.
Thus, metastability excludes parameter ranges beyond 
existing direct searches.

  Note as well that metastability does not place a lower bound on the
mass of the lightest stop.  For example, a very light state can be
obtained for $m_{U_3}^2 \ll m_{Q_3}^2$ and $X_t = 0$.  This scenario is
not constrained by metastability, and can generate
a SM-like Higgs boson mass consistent with observations for sufficiently
large values of $m_{Q_3}^2$~\cite{Carena:2008rt}.\footnote{
A lower limit on the light stop mass in this scenario
can be obtained from its effect on Higgs production and 
decay rates~\cite{Fan:2014txa}.}

  Our results also have implications for future stop searches
and measurements.  Should a pair of stops be discovered, a variety of methods
can be used to determine the underlying parameters in the stop mass
matrix through precision measurements at the 
LHC~\cite{Hisano:2002xq,Rolbiecki:2009hk,Blanke:2010cm,Perelstein:2012fc}
or a future $e^+e^-$ 
collider~\cite{Berggren:1999ss,Boos:2003vf,Perelstein:2012fc}.
If these stop parameters turn out to lie within the dangerously unstable region,
corresponding to larger values of $|X_t|$,
we can conclude that new physics beyond the MSSM must be present.

\subsection{Stop Bound States}

  An additional phenomenon that can potentially occur in the MSSM 
when $A_t$ is very large is the formation of a $\stp_L\stp_R^*$ bound 
state through the exchange of light Higgs 
bosons~\cite{Giudice:1998dj,Hernandez:2001mi}.  Such a state could have
the quantum numbers of a Higgs field and mix with the MSSM Higgs 
fields to participate in electroweak symmetry breaking~\cite{
Giudice:1998dj,Hernandez:2001mi,Cornwall:2012ea,Pearce:2013yja}.  
If this occurs, our results on the metastability of the MSSM may 
no longer apply.  Calculating the critical value of $A_t$ for when 
a bound state arises is very challenging, but under a set of 
reasonable approximations Ref.~\cite{Cornwall:2012ea} finds that 
it requires $A_t/M_S \gtrsim 15$.  While this lies beyond the region 
considered in the present work, it is conceivable that 
a full numerical analysis would yield a lower critical value for this ratio.

\section{Conclusions~\label{sec:conc}} 

  In this work we have investigated the limits on the stop parameter 
space imposed by vacuum stability considerations. A SM-like Higgs boson 
with a mass of $\sim 125\;\gev$ in the MSSM points to a particular 
region of the parameter space if naturalness of the EW scale is desired. 
In this regime, the two stop gauge eigenstates are highly mixed, 
and this can induce the appearance of charge- and color-breaking 
minima in the scalar potential. Quantum tunnelling to these vacua 
can destabilize the electroweak ground-state. 

  We have studied the conditions under which stop mixing can induce
CCB vacua and we have computed the corresponding 
tunnelling rates.  We find that metastability provides an important 
constraint on highly-mixed stops.  We have also considered constraints 
from flavor and precision electroweak observables and direct stop searches, 
which are sensitive to a similar region of the MSSM parameter space.  
Metastability provides new and complimentary limits, with a different
dependence on the underlying parameter values. 

  The metastability limits we have derived provide a necessary condition
on the MSSM.  They apply for both standard and non-standard cosmological
histories.  Let us emphasize, however, that the MSSM parameter points that we
have found to be consistent with stop-induced CCB limits may still be
ruled out by more general stability considerations, such as configurations
with more non-zero scalar fields.  Fortunately, our own SML vacuum
appears to be at least safely metastable.

\begin{flushleft}
\textbf{Note Added:}
  While this manuscript was in preparation, two other works
considering limits on the MSSM from vacuum stability 
appeared~\cite{Camargo-Molina:2013sta,Chowdhury:2013dka}.
A preliminary version of the present results was also posted
as a contribution to a conference proceedings~\cite{Blinov:2013uda}.
We study a different region of MSSM parameter space than 
Ref.~\cite{Camargo-Molina:2013sta}, but we do have significant overlap
with Ref.~\cite{Chowdhury:2013dka}.  Our results are substantially 
in agreement with Ref.~\cite{Chowdhury:2013dka}, although our exclusions 
from metastability extend to larger values of $X_t$.  We suspect that this 
difference is due to slightly different choices of parameters as well 
as variations in the outputs of the spectrum generators that were used 
in our respective analyses.
\end{flushleft}

\acknowledgments
We thank Aaron Pierce for the discussion that initiated this work. 
NB thanks Max Wainwright, Nikolay Blinov and Rishi Sharma for useful 
conversations. This research is supported by National Science 
and Engineering Research Council of Canada~(NSERC).

\newpage

\begin{appendix}

\section{Minimality of the Action Under Path Deformations\label{sec:stuff}}

  In this appendix we show that fixing a path in field space connecting
two vacua and computing the one-dimensional bounce action along that
path provides on upper bound on the bounce action for tunnelling
between those vacua.  Equivalently, the bounce solution of the Euclidean
action is a minimum of the action with respect to deformations of
fixed, one-dimensional paths in the field space.  The implication of
this result is that the path deformation method of 
CosmoTransitions~(CT)~\cite{Wainwright:2011kj} is guaranteed to provide 
at least an upper bound on the tunnelling lifetime.

  Recall that the multi-field bounce solution $\bvec{\phi}(\rho)$, 
$\rho=\sqrt{t^2+\vec{x}^2}$, of the Euclidean action is an 
$\mathcal{O}(4)$-symmetric solution of the classical equations 
of motion subject to the boundary conditions
\beq
\del_{\rho}\bvec{\phi}(\rho=0)=0 \ ,~~~~~
\lim_{\rho\to\infty}{\bvec{\phi}} = \vec{\phi}_+ \ ,
\eeq
where $\vec{\phi}_+$ is the metastable vacuum configuration.
The bounce action is just the Euclidean action evaluated on
the bounce solution.

  Let us now restate our claim more precisely.  
The bounce solution is an element of the set of parametric curves
on $\mathbb{R}^{F}$, where $F$ is the number of scalar fields. 
Any path $\vec\phi(\rho)$ in this set can be written in terms
of a unit speed curve $\vec{\gamma}(s)$:
\beq
\vec\phi(\rho) = \vec\gamma(s(\rho)) \ ,
~~~\text{where}~~~\left|\dot{\vec\gamma}(s)\right|=1 \ .
\eeq  
The function $s(\rho)$ is the solution of 
\beq
\frac{ds}{d\rho} = \left|\frac{d\vec\phi}{d\rho}\right| \ ,
\eeq 
and $\vec\gamma(s) = \vec\phi(\rho(s))$. 
The Euclidean action in $\alpha$ spacetime dimensions becomes
\beq
S_E[\vec\gamma,s] = \Omega_\alpha \int d\rho \rho^{\alpha-1}
\left[
\frac{1}{2}\left(\frac{ds}{d\rho}\right)^2
+ V(\vec\gamma(s(\rho)))
\right],
\label{eq:transformedaction}
\eeq
where $\Omega_\alpha = 2\pi^{\alpha/2}/\Gamma(\frac{\alpha}{2})$
is the surface area of a unit $(\alpha-1)$-sphere.
Suppose we fix a path in field space $\vec\gamma$ connecting two
vacua and extremize the action with respect to $s(\rho)$ subject to 
the boundary conditions of the bounce along this one-dimensional trajectory. 
The corresponding solution can then be used to obtain a restricted
bounce action along the fixed trajectory.  This is the procedure used by CT 
at each intermediate step of its deformation procedure.  We claim that
the action obtained for any such fixed path is greater than or equal
to the unconstrained bounce action.

  To prove this claim, we use the fact that the bounce is a stationary
point of the action.  For tunnelling configurations, however, it is
not an extremum of the action.  This coincides with the fact that
the second variation of the action with respect to the fields
has a negative eigenvalue.  The corresponding operator is
\beq
-\delta_{ij}\partial^2 + \frac{\delta V}{\delta \phi_i\delta \phi_j}(\bar{\vec\phi}) \ .
\eeq
We assume that this operator has only a single 
negative mode~\cite{Callan:1977pt,Claudson:1983et}. 
This has been proved for a single field in the thin wall 
limit~\cite{Callan:1977pt}. If this assumption is false, the entire 
Callan-Coleman formalism does not apply. 
We show that this negative eigenvalue is associated exclusively with 
the variation of $s(\rho)$ using the argument of Ref.~\cite{Claudson:1983et}. 
As a result, the bounce action is an extremum with respect to variations in the
orthogonal parameter $\vec\gamma$, and can easily be shown to be a minimum
by explicit construction.

Consider the scaling transformation
\beq
s(\rho)\rightarrow s(\rho/\lambda). 
\label{eq:scalingtform}
\eeq 
The action of Eq.~\eqref{eq:transformedaction} transforms as 
\beq
S[\vec\gamma,s]\rightarrow\lambda^{\alpha-2} S_T[\vec\gamma,s] 
+ \lambda^\alpha S_V[\vec\gamma,s],
\eeq
where 
\beq
S_T[\vec\gamma,s] = \Omega_\alpha\int d\rho \rho^{\alpha-1} 
\frac{1}{2}\left(\frac{ds}{d\rho}\right)^2
\eeq
and
\beq
S_V[\vec\gamma,s] = \Omega_\alpha\int d\rho \rho^{\alpha-1} 
V(\vec\gamma(s(\rho))).
\eeq
Requiring that $S$ is stationary with respect to these scale variations 
yields
\beq
\frac{\delta S}{\delta \lambda} = 0\Rightarrow S_T = -\frac{\alpha}{\alpha-2} S_V > 0. 
\eeq
We can also evaluate the second variation of $S$
\beq
\frac{\delta^2 S}{\delta \lambda^2} = 
\begin{cases}
-S_T & \alpha = 3 \\
-2(\alpha-2)S_T& \alpha > 3
\end{cases}
< 0. 
\eeq
This means that the bounce is a maximum of the action with respect 
to the scaling transformation of Eq.~\eqref{eq:scalingtform}. 
Thus the crucial negative eigenmode is due to scaling, and, 
since this transformation does not involve the normalized path $\vec\gamma$, 
it is due entirely to the functional variation of $s(\rho)$.
The tunnelling action obtained by computing the bounce solution along
a fixed one-dimensional path is therefore an upper bound on the true bounce
action.  This justifies the procedure of using a fixed normalized field path 
and computing $s(\rho)$ as a way to check the CosmoTransitions results.

\section{An Approximate Empirical Bound\label{sec:empbound}}

  In this second appendix we describe an approximate
empirical bound on metastability valid in the parameter region
$r\equiv m_{U_3}^2/m_{Q_3}^2 \sim 1$, moderate $\tan\beta$, smaller $\mu$,
and larger $m_A$.  We begin by deriving a condition on 
\emph{absolute} stability to motivate the functional form of 
the empirical formula.  Let us emphasize that our empirical bound
is only an approximation, and is not guaranteed to work outside 
the limited regime we consider.

  To derive an improved bound on absolute stability of the SM-like~(SML) vacuum,
we impose only $SU(3)_C$ $D$-flatness and $H_d^0 = 0$. Similar 
existing formulae typically also assume $SU(2)_L$ and $U(1)_Y$ flatness, 
which precludes the existence of a SML vacuum. 
For $m_{U_3}^2/m_{Q_3}^2 \sim 1$, $SU(3)_C$ $D$-flatness should be a good
approximation since the strong gauge coupling is larger than 
the others~\cite{LeMouel:2001sf}. 
Setting $H_d^0 = 0$ is also well-justified for large $\tan\beta$ near the SML 
vacuum; at the CCB minimum one typically finds $|H_d^0| < |H_u^0|$ as well.

Applying the $SU(3)_C$ $D$-flatness condition, we have
\beq
T \equiv\stp_L = |\stp_R| \ ,
\eeq
and the potential becomes 
\beq
V &=& m_T^2T^2 + m_2^2(H_u^0)^2 
\pm 2y_tA_tH_u^0T^2
%\\&&\hspace{0.5cm}
+ y_t^2\left[T^4+ 2T^2(H_u^0)^2\right]
+ \frac{\bar{g}^2}{8}\left[(H_u^0)^2- T^2\right]^2 \ ,\nnmb
\eeq
where $m_T^2 = m_{Q_3}^2+m_{U_3}^2$, $\bar{g} = \sqrt{g^2+{g'}^2}$ 
and $m_2^2 = m_{H_u}^2 + |\mu|^2$. 

 Minimizing, we have
\beq
0 = \frac{\del V}{\del T} 
= T\left[2m_T^2\pm 4y_tA_tH_u + 4y_t^2H_u^2 
- \frac{\bar{g}^2}{2}((H_u^0)^2-T^2) + 4y_t^2T^2\right] \ .
\eeq
The solutions are evidently $T=0$ and 
\beq
T^2 = \left[\mp 2y_tA_tH_u^0 - m_T^2 - 2(y_t^2-\bar{g}^2/8)(H_u^0)^2\right]
\bigg{/}
2(y_t^2+\bar{g}^2/8) \ .
\eeq
Since we are restricting ourselves to $H_u^0\geq 0$, the relative orientation
of the stops in any potential CCB minimum must be such that 
$\mp y_tA_t = |y_tA_t|$.  Note as well that the $A$-term must overpower
the others to make $T^2>0$.  Under our given assumptions, this already
provides a necessary condition on the existence of a CCB vacuum,
\beq
A_t^2 > 2m_T^2(1-\bar{g}^2/8y_t^2) \ .
\eeq
This is a somewhat weaker requirement than the analytical formula 
Eq.~\eqref{eq:analyticalbound}.

  Minimizing with respect to $H_u$ (and choosing the relative stop
alignment as above) gives
\beq
0 = \frac{\del V}{\del H_u} = 
2m_2^2H_u^0 + 4(\bar{g}^2/8)(H_u^0)^3 
+ \left[(2(y_t^2-\bar{g}^2/8)H_u^0-y_tA_t\right]\,(2T^2) \ .
\eeq
For $T^2=0$, this reproduces the SM-like minimum.  
On the other hand, we can also plug in our non-zero
solution for $T^2$, which is quadratic in $H_u^0$.  This generates
a cubic equation for $H_u^0$ that can be solved analytically.
A cubic equation has three roots, with at least one of them real.
The other two roots are either real, or complex conjugates of each other.
We need at least three real roots to have both a SML vacuum and 
a CCB vacuum since there must also be at least a saddle point between them.

In this approximation, we can check for CCB vacua by simply scanning over
stop parameters and computing cube roots, for which there exist analytical
formulae. The EW vacuum is trivial to find, and corresponds to $T=0$.
The $T^2\neq 0$ solutions may correspond to CCB vacua.  A necessary condition
for this is that all the roots are real, and that at least two of them are
positive.  With the roots in hand, it is then straightforward to use them
in the potential to compare the relative depths of the minima.
Fixing $m_2^2 = -m_Z^2/2$ to get the correct SML vacuum 
expectation value, we find numerically that $A_t^2 \gtrsim 
(2.4)(m_T^2+{m}_{2}^2)$ gives a very good estimate of the condition 
for a CCB vacuum to be deeper than the SML vacuum for this simplified potential. 

  In our analysis of metastability, we find that the boundary
between metastable and dangerously unstable regions tends to track
the boundary between SML and CCB regions.  Motivated by this and
our previous result for CCB vacua, we will attempt to fit the boundary
between metastable and unstable regions by an expression of the form
\beq
A_t^2 = \alpha m_T^2 + \beta |m_{Q_3}^2 - m_{U_3}^2| + \gamma m_2^2 
 = \left(
\alpha + \beta \frac{|1-r|}{1+r}
\right)m_T^2 + \gamma m_2^2 
\label{eq:fitform}
\eeq 
The second term in the above expression is included 
to model the effect of small deviations from $SU(3)_C$ $D$-flatness. 

\begin{figure}[ttt]
 \begin{center}
         \includegraphics[width = 0.47\textwidth]{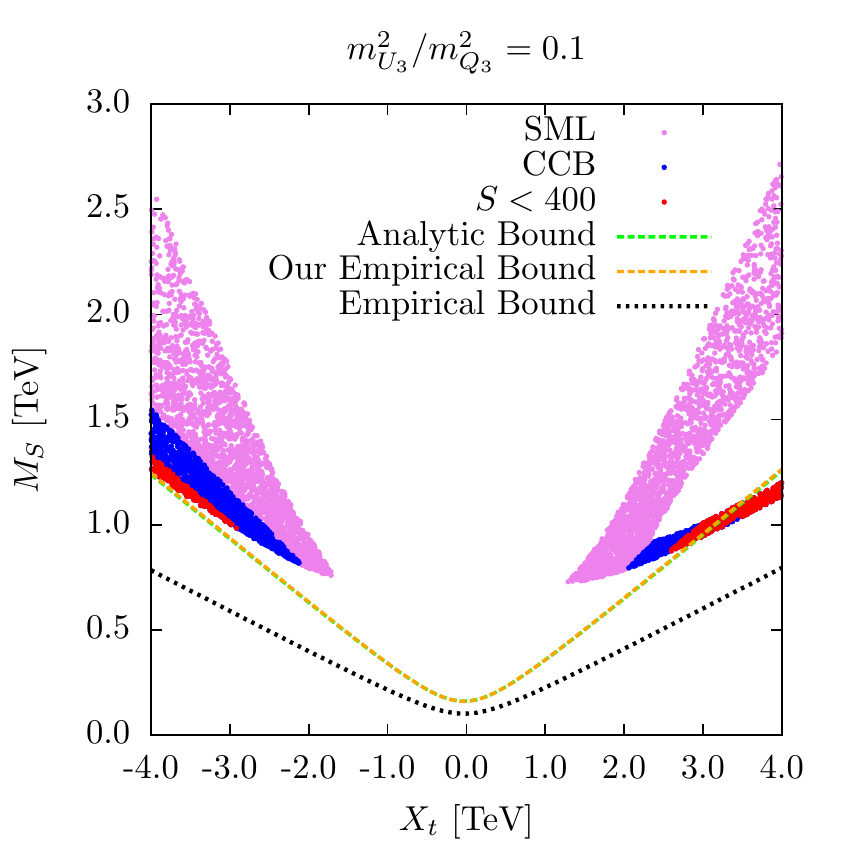}
         \includegraphics[width = 0.47\textwidth]{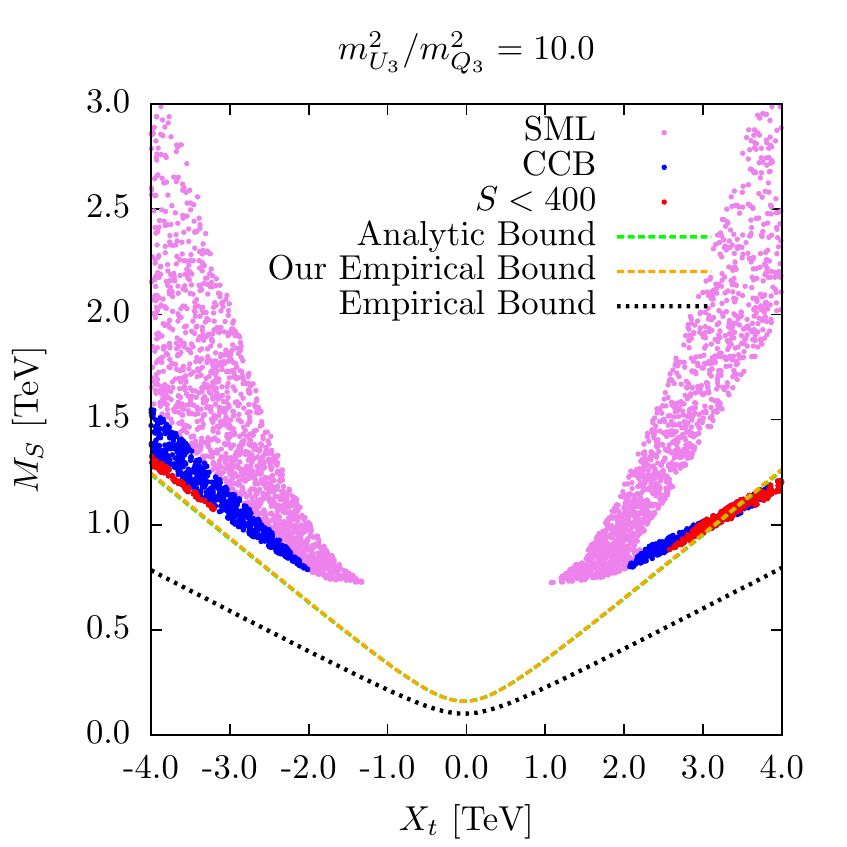}
 \end{center}
 \caption{The deterioration of the empirical bound of Eq.~\eqref{eq:fitform} for 
 $m_{U_3}^2/m_{Q_3}^2 \gg 1$ or $m_{U_3}^2/m_{Q_3}^2 \ll 1$. For these parameter ranges 
the assumption of $SU(3)_C$ $D$-flatness that motivated Eq.~\eqref{eq:fitform} breaks down 
and it cannot be used to reliably model the boundary between the metastable and unstable 
parameter regions.
 \label{fig:extremeratios}}
 \end{figure}

  We use estimate the parameters $\alpha$ and $\gamma$ by using a least-squares 
fit to the lower boundary of the metastable region in 
Fig.~\ref{fig:nomh_default}, without imposing the Higgs mass constraint. 
This is an arbitrary choice to fit to; different choices 
in Tab.~\ref{tab:scanrange} lead to variations in $\alpha$ on the order 
of $15\%$ and $100\%$ in $\gamma$. The large variation in $\gamma$ 
is not a big problem since it is multiplied by $|m_2^2| \sim m_Z^2 \ll m_T^2$. 
We obtain $\alpha\simeq 3.4$ and $\gamma\simeq 60$. 
With $\alpha$ and $\gamma$ fixed, we fit $\beta$ to models with 
$r\neq 1$. We again see that there is a significant variation 
$\mathcal{O}(20\%)$ depending on what $r$ is, indicating that 
the functional form of Eq.~\eqref{eq:fitform} is an oversimplification. 
With this in mind, we find an average value of $\beta\simeq 0.5$ 
for $r\in [0.3,3]$.  We show the resulting bound in the results 
of Section~\ref{sec:results}. For $r\simeq 1$, Eq.~\eqref{eq:fitform} 
approximates the true boundary between metastable and unstable models well. 
However, we expect this constraint to deteriorate as one moves away 
from the assumption of $SU(3)_C$ $D$-flatness by choosing  soft 
masses with $r\gg 1$ or $r \ll 1$. We show an explicit example of
this in Fig.~\ref{fig:extremeratios}.

We emphasize that this bound is a \emph{very} rough guideline 
for metastability in the MSSM in a specific corner of the parameter 
space and should only be used as a first order approximation. 
A full numerical analysis is required when any of the above assumptions 
are violated or better precision is required.

\end{appendix}

\end{document}